\documentclass[reprint,numberedrefs,authaffil]{JASAnew2}

\usepackage{bm}
\newcommand{\ve}[1]{\bm{#1}}

\newcommand{\bmm}{\ve{m}}

\newcommand{\bu}{\ve{u}}

\newcommand{\bp}{\ve{p}}

\newcommand{\bq}{\ve{q}}

\newcommand{\bR}{\ve{R}}

\newcommand{\bF}{\ve{F}}
\newcommand{\bI}{\ve{I}}

\newcommand{\bphi}{\boldsymbol{\phi}}

\newcommand\SPL{\mathrm{SPL}_\mathrm{max}} 
\newcommand\SPLa{\mathrm{SPL}_\mathrm{avg}}

\newcommand{\overbar}[1]{\mkern 1.5mu\overline{\mkern-1.5mu#1\mkern-1.5mu}\mkern 1.5mu}

\usepackage{amsmath}
\usepackage{amssymb}
\usepackage{float}
\usepackage[mediumspace,mediumqspace,squaren]{SIunits}

\newcommand{\species}[1]{\emph{#1}}

\begin{document}

\title{
    Simulation of humpback whale bubble-net feeding models
}

\author{Spencer H. Bryngelson}
\email{spencer@caltech.edu}
\affiliation{Division of Engineering and Applied Science, 
    California Institute of Technology,
    1200 E California Blvd, Pasadena, CA 91125, USA}

\author{Tim Colonius}
\affiliation{Division of Engineering and Applied Science, 
    California Institute of Technology,
    1200 E California Blvd, Pasadena, CA 91125, USA}

\begin{abstract}
    
    Humpback whales can generate intricate bubbly regions, called bubble nets, via their blowholes.
    They appear to exploit these bubble nets for feeding via loud vocalizations.
    A fully-coupled phase-averaging approach is used to model the flow, bubble dynamics, and corresponding acoustics.
    A previously hypothesized waveguiding mechanism is assessed for varying acoustic frequencies and net void fractions.
    Reflections within the bubbly region result in observable waveguiding for only a small range of flow parameters.
    A configuration of multiple whales surrounding and vocalizing towards an annular bubble net is also analyzed.    
    For a range of flow parameters the bubble net keeps its core region substantially quieter than the exterior.
    This approach appears more viable, though it relies upon the cooperation of multiple whales.
    A spiral bubble net configuration that circumvents this requirement is also investigated.
    The acoustic wave behaviors in the spiral interior vary qualitatively with the vocalization frequency and net void fraction.
    The competing effects of vocalization guiding and acoustic attenuation are quantified.
    Low void fraction cases allow low-frequency waves to partially escape the spiral region, with the remaining vocalizations still exciting the net interior.
    Higher void fraction nets appear preferable, guiding even low-frequency vocalizations while still maintaining a quiet net interior.

\end{abstract}

\maketitle

\section{Introduction}\label{s:intro}

\begin{figure*}[t]
	\centering
    \includegraphics[scale=1]{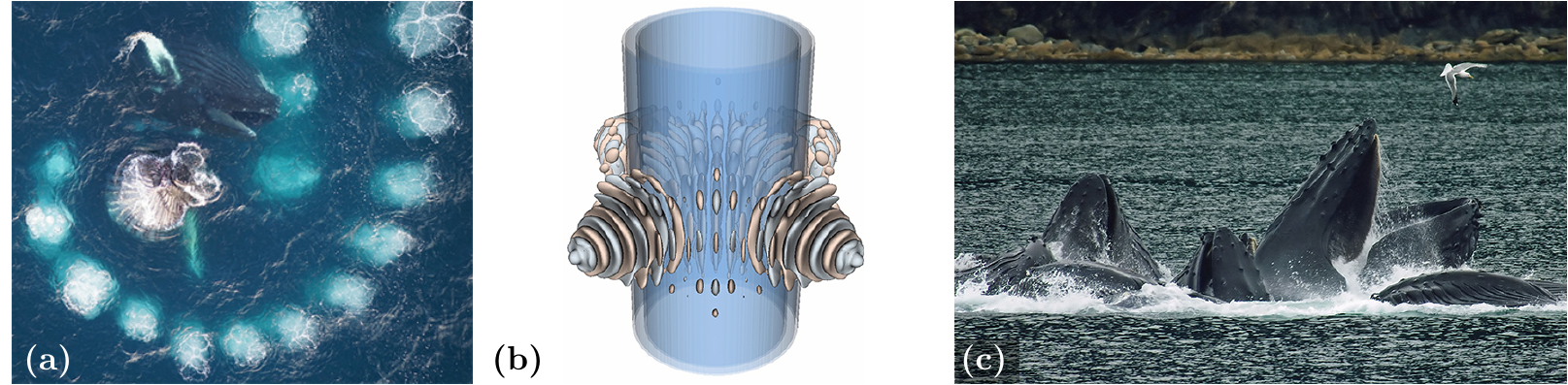}
	\caption{%
	(a) Aerial view of a humpback whale bubble net~\citep{whale_img2},
	(b) visualization of four whales vocalizing towards a bubble net 
	using the present simulation method (see section~\ref{s:model}), and
	(c) several humpback whales lunge feeding~\citep{whale_img1}.
	}
	\label{f:whales}
\end{figure*}

Humpback whales (\textit{Megaptera novaeangliae}) utilize sophisticated underwater feeding strategies~\citep{ingebrigtsen29}.
They can generate bubbles with their dorsal surface (blowholes) and form bubble columns~\citep{hain81}, clouds~\citep{hain81}, and nets~\citep{jurasz79} with complex swimming maneuvers~\citep{fish95}.
The whales appear to leverage these bubbly regions via acoustic excitation for trapping and corralling small fish (mostly herring and krill~\citep{hain81}).
Indeed, the whale vocalization frequencies often even overlap with the resonant frequencies of the fish swim bladders~\citep{leighton07c,leighton04c}.
However, the mechanisms by which the whales exploit (or suffer from~\citep{leighton14}) these nets are generally unknown.
The bubble-net feeding strategy is focused on here, for which the whales swim downwards in a circular motion, starting from a few meters below the ocean surface.
They then rotate their blowholes towards the will-be bubble-net center and release several ``bursts'' of bubbles~\citep{hain81}, creating an annular or spiral cylinder of bubbles~\citep{sharpe97} as shown in figure~\ref{f:whales}~(a) and (b)
From the net exterior, from one to about 15~\citep{valsecchi02} whales generate a loud trumpeting sound.
The whales then rise within the net center and consume their prey in a process called vertical lunge feeding~\citep{dvincent85} (see figure~\ref{f:whales}~(c)).

Observations of bubble nets are scarce, as they usually occur in isolated ocean areas.
As a result, hypotheses of how whale generated acoustic waves and their interactions with bubble nets result in advantageous feeding strategies are broad.
For example, it is possible that whales utilize their nets to echolocate prey~\citep{mercado01}, though this could require higher frequencies than those usually observed during the bubble-net feeding process.
Here, the possibility that the whales attempt to surround and trap their prey with loud vocalizations is considered for different acoustic mechanisms.

The first mechanism considered follows from~\citet{leighton04}, who showed that the bubbly region could behave as a waveguide for the whale vocalizations if certain criteria are met.
For this, the acoustic waves would enter (via refraction) and reflect within an annular bubbly region, eventually occupying the entire bubble net and trapping the prey within a wall of sound.
However, for this to be possible, the vocalizations must be focused in a sufficiently narrow angular range, the bubble natural frequency must be higher than the vocalization frequency, and the sound must remain sufficiently directional within the net~\citep{leighton04_2}.
While these criteria are plausible, it is unknown if geometric effects or nonlinear and collective bubble interactions or oscillations will preclude a useful degree of waveguiding.
Indeed, \citet{leighton07c} later hypothesized that it is improbable these criteria could be reliably and simultaneously met, though it remains unclear if this mechanism is physically viable for reasonable model parameterizations.

If the net geometry is annular, then it is possible that the bubble net simply shields the interior from the vocalizations, which then surround the net region~\citep{leighton04c}.
Bubble curtains have been used to shield krill from damage in captivity~\citep{finley03} and herring are reluctant to cross high void fraction bubble curtains even in the absence of acoustic excitation~\citep{sharpe97}.
Thus, if the effective impedance of the bubble net is large and acoustic refraction is insignificant, then this conjecture is plausible.
However, a ready assessment of this appears to be precluded by the non-uniform void fraction within the net, acoustic interactions between multiple whales, and geometric effects of the circular bubbly region.

More recently, it was noticed that the nets might instead have spiral shapes~\citep{leighton07_2}.
Other acoustic mechanisms are possible if this is the case.
For example, \citet{leighton07} described that the whales could reflect their vocalizations within the bubble-free region of the net, which might allow a single whale to surround the entire net region with sound.
Further, this would utilize a larger fraction of the energy they generate, instead of squandering the portion that is reflected away from the net in the waveguide scenario~\citep{leighton07c}.
However, like above, it is challenging to anticipate and confirm a full operating mechanism via only theoretical and ray-tracing analysis due to the non-uniform bubbly regions, collective bubbly effects, low-frequency behaviors, and the sum-and-difference frequencies that could arise.

As a step towards understanding this feeding strategy, the present goals are to determine if a waveguiding behavior can be observed, to what degree the interior of the bubble net is kept quiet when directly excited by loud vocalizations, and the acoustic behavior and attenuation for spiral net configurations.
It is possible to perform \textit{in situ} or laboratory experiments to analyze such configurations.
For example,~\citet{leighton07_2} used expanded polystyrene to model the acoustic impedance.
However, it remains challenging to reliably control the bubble population distribution or accurately observe the reflection and refractions that occurs near the bubble wall.
This motivates the use numerical simulations to consider a range of possible flow parameters (net void fraction, vocalization frequency and orientation, etc.) and net geometries.
Here, the simulation model used includes bubble--bubble interactions, nonlinear bubble dynamics including surface tension, viscosity, and mass transfer, geometric effects due to the bubble-net wall thickness, and the finite breadth of the driving acoustics.
This model and the numerical methods used for its solution are described in greater detail in section~\ref{s:model}.
Results are presented in section~\ref{s:results} for acoustically excited circular and spiral bubble nets for a range of model and flow parameters.
The implications of these results are discussed in section~\ref{s:conclusion}.

\section{Model system and numerical methods}\label{s:model}

\subsection{Problem setup}\label{s:setup}

\begin{figure*}
	\centering
	\includegraphics[scale=1]{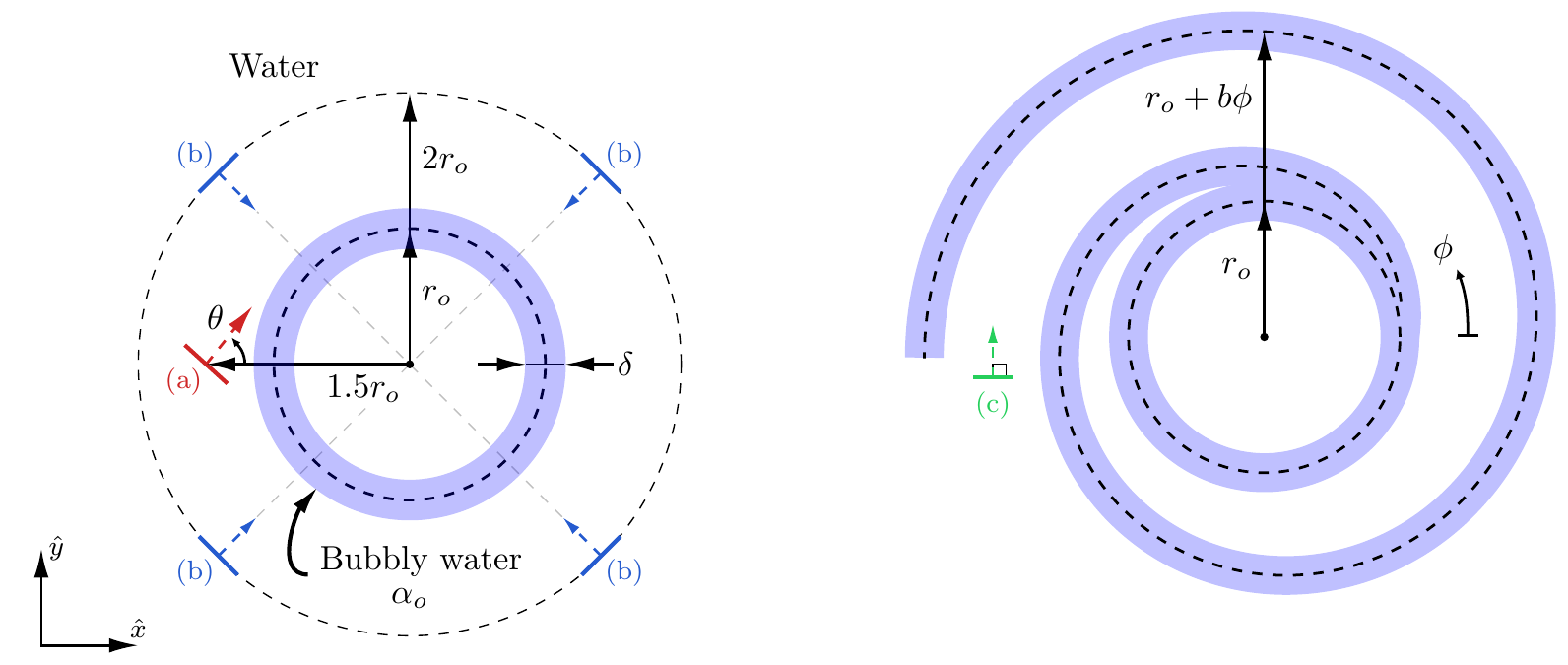}
	\caption{%
    Problem setups: Annular (left) and spiral bubble-nets (right).
    The lines (a--c) indicate different acoustic source locations and directions (see text).
	}
	\label{f:setup}
\end{figure*}

Schematics of the problems considered are shown in figure~\ref{f:setup}.
The nominal height of the bubble net is at least as large as its nominal radius~\citep{wiley11}, so the bubble net flow system is modeled as two dimensional on average, though the individual bubbles are spherical.
This model does not account for some of the more intricate upward spiral or double-loop three-dimensional bubble net geometries that have been reported~\citep{wiley11}, nor buoyant effects due to rising bubbles.
Instead, it is utilized to test the hypotheses discussed in section~\ref{s:intro}.
The net region is either an annulus of radius $r_o$ and nominal thickness $\delta$ (figure~\ref{f:setup}, left) or the same annulus with the addition of an Archimedean spiral of radius $r = r_o + b \phi$, turning parameter $b = r_o/\pi$, and the same thickness (figure~\ref{f:setup}, right).
Net parameters $r_o = \unit{10}{\meter}$ and $\delta = \unit{4}{\meter}$ are used, both of which are reasonable estimates for actual bubble nets~\citep{dvincent85}.
The net is located at the center of a side-length $L = 6 r_o$ square domain.
It is filled with spherical gas bubbles of density $\rho_g = \unit{1}{\kilo\gram\per\meter\cubed}$ and radius $R_o = \unit{1}{\milli\meter}$, which follows from the approxmimation of~\citet{wiley11}, though variations of these parameters will be discussed.
The bubbly water has void fraction $\alpha$ that varies with radial coordinate $r$ up to its initial maximum value $\alpha_o$ at $r= r_o$ via a Gaussian bump as
\begin{gather} \alpha (r)=
    \alpha_o \exp\left( -\frac{1}{2} \frac{(r - r_o)^2}{\sigma_\delta^2} \right),
\end{gather} 
where $\sigma_\delta = \delta/3$. 
The domain is otherwise occupied with water of density $\rho_l = \unit{998}{\kilo\gram\per\meter\cubed}$.

Whale vocalizations are represented as one-way waves emitted from a line-source of length $L_a = 0.2 r_o$ at a specified angle $\theta$, frequency $f$, and amplitude $A$.
Humpback whales can produce sounds ranging from \unit{10}{\hertz} to \unit{30}{\kilo\hertz}~\citep{mercado99,mercado01}, though the trumpeting calls associated with bubble net feeding are usually around a few kilohertz~\citep{thompson86}.
Constructive sum-and-difference frequencies could also extend the effective frequency observed at the bubble-net wall~\citep{leighton04}.
Thus, a frequency range of $f = \unit{0.1}{\kilo\hertz}$ to $\unit{5}{\kilo\hertz}$ will be discussed herein.
For the bubble sizes considered, this corresponds to both attenuated and enhanced effective mixture sound speeds in the bubbly region~\citep{commander89}.
The sound pressure of the vocalizations is about $A = \unit{180}{\deci\bel}$ re $\unit{1}{\micro\pascal}$~\citep{thompson86}, which is used here.
These sounds usually last for minutes~\citep{dvincent85}, which is much longer than the single-bubble-oscillation time scales
Thus, the acoustics-generating sources are active for the duration of the simulations.

The locations of the acoustic line sources are shown in figure~\ref{f:setup}.
The first configuration, shown in figure~\ref{f:setup}~(a), is a single line source with angle $\theta = 50^\circ$ from the $\hat{x}$-direction, which is used to determine if a waveguiding behavior can be observed in the bubbly region.
The second configuration (figure~\ref{f:setup}~(b)) is $N_w$ line sources ($N_w = 4$ shown), each $2 r_o$ from and directed towards the bubble net center, which is used to assess the effective acoustic impedance of the bubble net.
The last configuration, shown in figure~\ref{f:setup}~(c), is a single line source of angle $\theta = 90^\circ$, directed into the bubble-free arm of a spiral bubble net. 
Note that these sources are fixed in space, and thus Doppler effects are not considered.
However, the broad frequency range considered, along with the relative insensitivity of the results near the ends of this range, confirms that such effects are insignificant for current purposes.

\subsection{Physical model}\label{s:phymodel}

The flow of a dilute suspension of bubbles in a compressible liquid is modeled using ensemble phase averaging~\citep{zhang94}.
This model is able to reproduce the correct bubbly-mixture sound speeds, nonlinear bubble dynamics, and their coupling to the suspending liquid~\citep{ando10,bryngelson19}.
The mixture-averaged equations of motion are written in quasi-conservative form~\citep{commander89}:
\begin{gather}
	\frac{\partial \bq }{\partial t } + \nabla \cdot \bF = \bm{0}
	\label{e:goveq}
\end{gather}
where $\bq = \left\{ \rho, \rho \bu, E \right\}$ are the conservative variables and $\bF = \left\{ \rho \bu, \rho \bu \bu + p \bI, (E + p) \bu \right\}$ are the fluxes.
Here, $\rho$, $\bu$, $p$, and $E$ are the mixture density, velocity vector, pressure, and total energy, respectively.
The mixture pressure is
\begin{gather}
	p = (1-\alpha)p_l +
	\alpha  \left(
		\frac{\overbar{\bR^3 \bp_{bw} }}{\overbar{ \bR^3}} - \rho \frac{ \overbar{ \bR^3 \dot{\bR}^2 }}{ \overbar{\bR^3} }
	\right),
\end{gather}
for which $\bR$, $\dot{\bR}$, and $\bp_{bw}$ are the radius, radial velocity, and wall pressure of the bubbles, respectively.
These quantities are vectors that depend upon the equilibrium bubble sizes $\bR_o$ as $\bR(\bR_o) = \{ R_1, R_2, \dots, R_{N_b} \}$, where $N_b = 31$ is the number of bins that describes the assumed log-normal distribution function of relative scale parameter $\sigma$~\citep{bryngelson19}.
Overbars $\overbar\cdot$ denote the usual moments with respect to this distribution.
Note that cases considered here are monodisperse unless stated otherwise, for which the bubble dynamic variables are scalars instead.

The liquid pressure $p_l$ follows from the stiffened-gas equation of state as parameterized by the specific heat ratio $\gamma_l$ and stiffness $\Pi_\infty$~\citep{menikoff89}.
The void fraction is transported as
\begin{gather}
	\frac{ \partial \alpha}{\partial t} + \bu \cdot \nabla \alpha =
	3 \alpha \frac{ \overbar{\bR^2 \dot{\bR} }}{ \overbar{\bR^3} },
\end{gather}
where the right-hand-side represents the change of averaged bubble volume.
The void fraction is transported as

The associated bubble dynamics are evaluated as
\begin{gather}
	\frac{ \partial n \bphi}{\partial t} + \nabla \cdot (n \bphi \bu) = n \dot\bphi,
\end{gather}
where $\bphi \equiv \left\{ \bR, \dot\bR, \bp_b, \bmm_v \right\}$ are the bubble dynamic variables, as will be described next, and $n$ is the bubble number density per unit volume
\begin{gather}
	n = \frac{3}{ 4 \pi} \frac{\alpha}{ \overbar{\bR^3} }.
\end{gather}

The bubbles as assumed to be spherical, ideal, and spatially uniform gaseous regions~\citep{bryngelson19}.
Their dynamics are driven by pressure fluctuations of the surrounding liquid and their radial velocities and accelerations are computed via the Keller--Miksis equation~\citep{keller80}
\begin{equation}
    \begin{split}
        R \ddot{R} \left( 1 - \frac{ \dot{R} }{c} \right) &+ \frac{3}{2} \dot{R}^2 \left( 1 - \frac{ \dot{R} }{3c} \right) = \\
    &\frac{ p_{bw} - p_\infty }{\rho} \left( 1 + \frac{ \dot{R} }{c} \right) + \frac{ R \dot{p}_{bw} }{\rho c},
\end{split}
\end{equation}
where $c$ is the sound speed, $p_\infty$ is the bubble forcing pressure, and
\begin{gather}
    p_{bw} = p_b - \frac { 4 \mu \dot{R} } { R } - \frac { 2 \sigma } { R }
\end{gather}
is the bubble wall pressure.
The internal bubble pressure $p_B$ and the mass of the bubble contents $m_v$ follow from a reduced model that can represent heat and mass transfer~\citep{preston07}.
In whole, this single-bubble model includes thermal effects, viscous and acoustic damping, and phase change.

\subsection{Numerical methods}\label{s:numerics}

The model problem of figure~\ref{f:setup} is spatially discretized via a rectilinear and uniformly spaced grid with $N = 3 \times 10^3$ grid points in each coordinate direction $\hat{x}$ and $\hat{y}$, and thus $\Delta_x = \Delta_y = L/N$ are the mesh spacings.
Non-reflective boundary conditions are used to minimize finite-$L$ effects, though the principal results were found to be insensitive to doubling $L$ and $N$.
The numerical scheme used has been described in detail before~\citep{coralic14} and is integrated into the MFC open-source solver~\citep{bryngelson19mfc}.
Thus, it is only briefly discussed here.
The fluxes of~\eqref{e:goveq} are split spatially and integrated within cell-centered finite volumes.
The primitive variables are reconstructed at the finite-volume-cell faces via a 5th-order WENO scheme~\citep{coralic14} and the HLLC approximate Riemann solver is used to compute the fluxes~\citep{toro94}.
The time derivative is computed using the 3rd-order TVD Runge--Kutta algorithm~\citep{gottlieb98} and the step size follows from the usual CFL criterion, which is fixed at $0.1$ based upon the speed of sound of water.

\section{Results}\label{s:results}

\subsection{Observation of wave guidance in an annular bubble net}\label{s:waveguide}

The possibility of a acoustic waveguiding behavior in the model bubble net is considered first.
Such wave guidance would entail a bending of incoming acoustic waves into the bubbly region from the exterior due to the change of sound speed, and subsequent reflection of the waves back into this region when they reach the net inner wall~\citep{leighton04}.
For this, the configuration of figure~\ref{f:setup}~(a), which represents a directional humpback whale vocalization grazing a bubble net, is simulated and then further parameter variations are discussed.

\begin{figure*}
    \centering
    \includegraphics[scale=1]{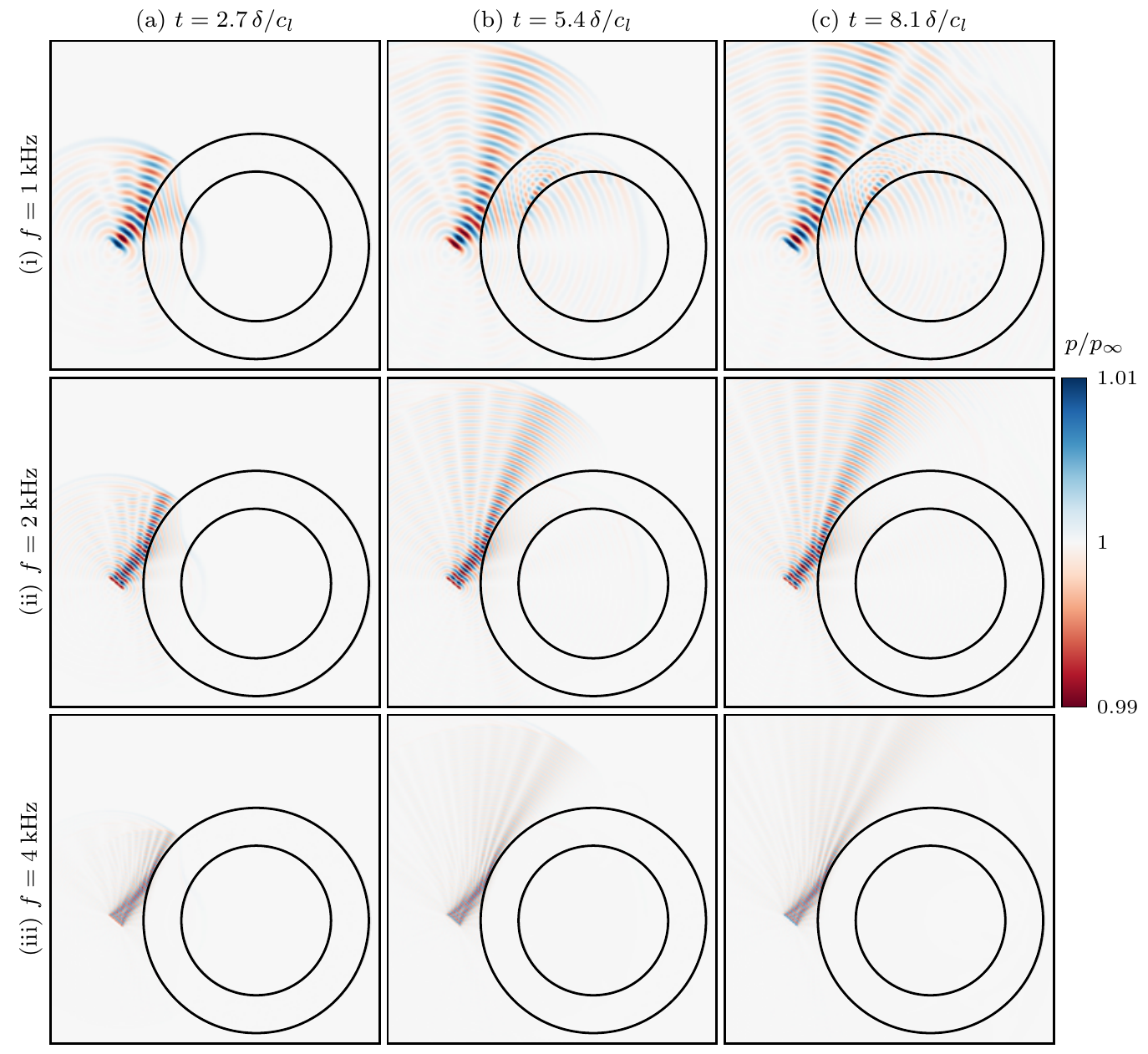}
    \caption{
        Pressure $p$ for bubble-net void fraction $\alpha_o = 10^{-4}$,
        (i)---(iii) varying acoustic source frequency $f$ and (a)---(c) times.
        The pressure scale corresponds to the maximum deviations from ambient pressure $p_\infty$ due to a $\unit{180}{\deci\bel}$ pulse.
        Times are non-dimensionalized by the net thickness $\delta$ and liquid sound speed $c_l$.
    }
    \label{f:viz_dirsrc}
\end{figure*}

Figure~\ref{f:viz_dirsrc} shows the time evolution of a directional acoustic wave of grazing a bubble net for varying vocalization frequencies.
For the lowest frequency considered $f = \unit{1}{\kilo\hertz}$, wave guidance is observed until $t = 8.1 \delta/c_l$.
This includes at least two reflections from the bubble net region that are clearly visible.
However, this behavior is nominally steady state and no further guidance is observed.
For $f = \unit{2}{\kilo\hertz}$ the vocalization refracts towards the net center at $t = 2.7 \delta/c_l$, though by $t = 5.4 \delta/c_l$ no further wave guidance is observed.
For the highest frequency $f = \unit{4}{\kilo\hertz}$ the effective impedance of the bubble wall is large and no waves are observed in the bubbly region, though it was confirmed that decreasing $\alpha_o$ until this impedance is small results in less wave-guidance than that observed for $f = \unit{1}{\kilo\hertz}$.
The lack of continual wave-guidance, as was most pronounced in the $f = \unit{1}{\kilo\hertz}$ case, appears to be due to the dispersion of the waves as they reflect in the non-uniform bubbly region.

\begin{figure*}
	\centering
    \includegraphics[scale=1]{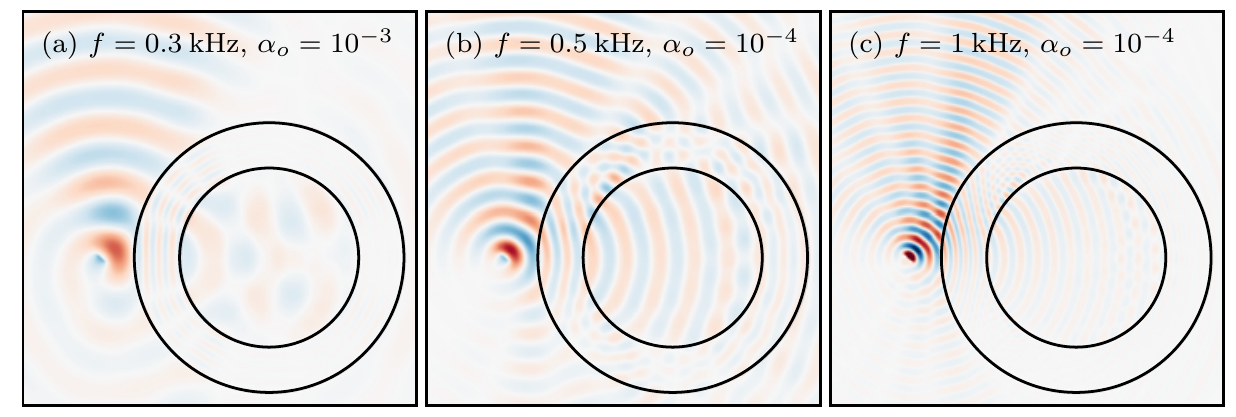}
	\caption{
        Example $\sigma = 0.7$ polydisperse cases with larger $\alpha_o$ and smaller $f$. 
        Normalized pressure contours are shown.
    }
	\label{f:viz_poly}
\end{figure*}

Considering monodisperse bubble populations means that, in order to optimally promote wave guidance, the resulting acoustic wave frequency must overlap with the single resonance frequency of the bubble population.
This challenge can be partially alleviated by considering polydisperse bubble populations with a large span of such resonant frequencies. 
This is considered in figure~\ref{f:viz_poly}, which uses a $\sigma = 0.7$ log-normally distribution of bubble sizes, as is considered typical for samples of sea water~\citep{ohern88}.
Further, lower frequencies and larger void fractions are also considered as a demonstration of waveguide potential
Figure~\ref{f:viz_poly}~(a) show a $f = \unit{0.3}{\kilo\hertz}$ case with larger $\alpha_o = 10^{-3}$ void fraction.
For this case the wavelength of the exciting frequency exceeds $\delta$, thus prohibiting wave-guidance. 
The $f = \unit{0.5}{\kilo\hertz}$, $\alpha_o = 10^{-4}$ case of figure~\ref{f:viz_poly}~(b) shows that further decreasing $f$ from the previous cases considered results in a significant pollution of the bubble-net center.
This, combined with the lack of additional waveguiding observed, suggests that such a configuration is less effective for this feeding strategy.
Finally, in figure~\ref{f:viz_poly}~(c) an otherwise previously considered case is shown, with the exception of the polydispersity and smaller acoustic source width of $L_a = 0.2 r_o$ (representing the possibility of a parametric sonar-like effect~\citep{leighton04_2}).
Again, modifying these parameters does not promote waveguiding, seemingly due to the significant dispersion observed in the bubbly region.
Further, varying the bubble-net thickness ($\delta$), grazing angle ($\theta$), and amplitude ($A$) of the incoming vocalizations did not observe wave-guidance beyond about two reflections from the bubble-net wall.

Thus, these results support prior claims that in practice the annular bubble net likely does not reliably act as a waveguide~\citep{leighton04c}.
Of course, additional whales carefully organized around the net and vocalizing in a similar fashion could promote excitation of the bubbly region.
However, if several whales are present and cooperating, it might be more likely that they are exploiting a more simple mechanism: they are utilizing the bubble net as an acoustic shield.
This possibility is investigated next.

\subsection{Acoustic shielding of annular bubble nets}\label{s:shield}

Multiple whales vocalizing towards a bubble net are modeled via the configuration of figure~\ref{f:setup}~(b).
The flow is simulated until a steady state is reached and the resulting acoustic waves are visualized and analyzed.

\begin{figure*}
	\centering
	\includegraphics[scale=1]{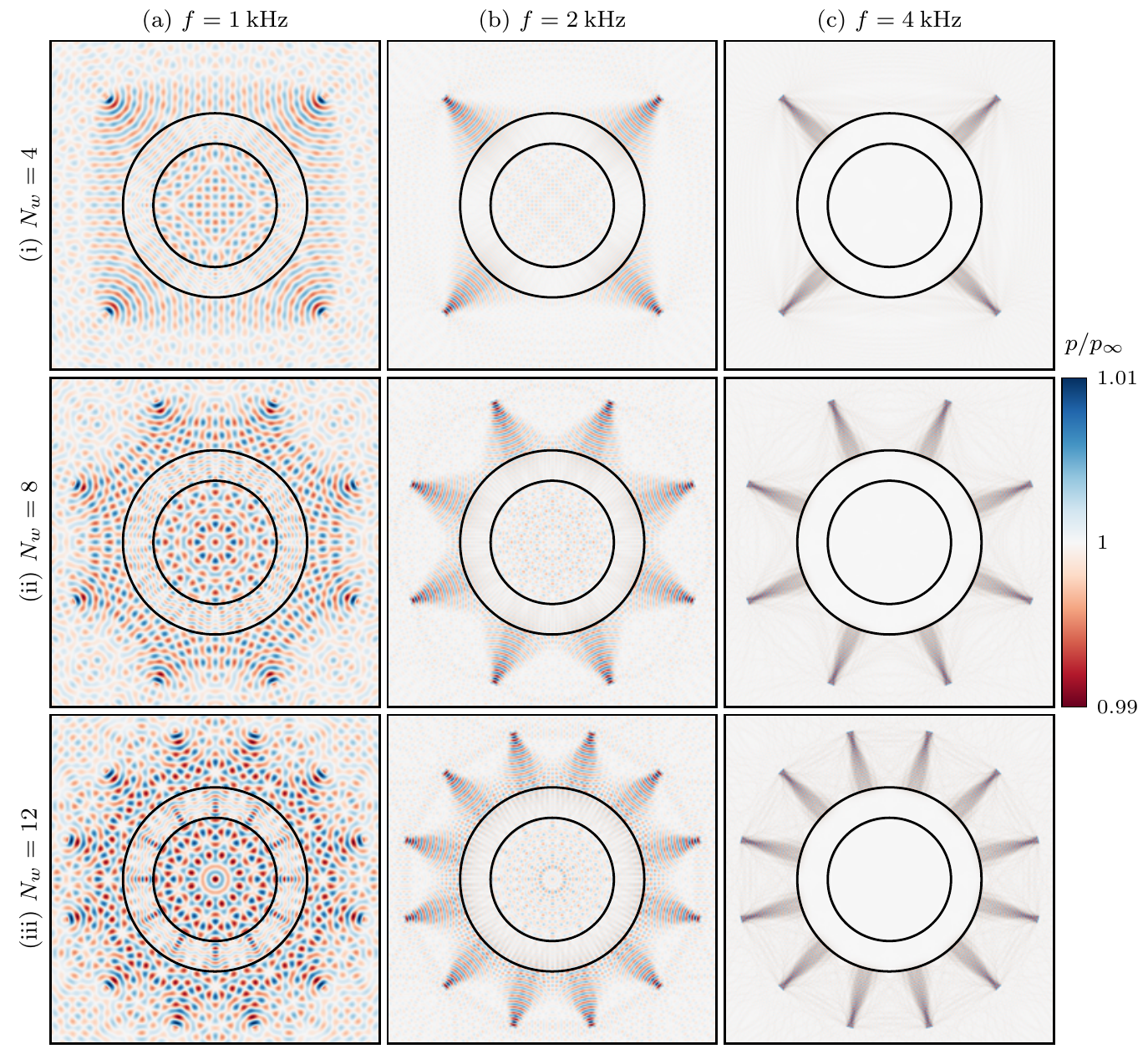}
	\caption{
	Long-time pressure $p$ for $\alpha_o = 10^{-4}$ and
	(a)--(c) varying acoustic source frequency $f$ and
	(i)--(iii) their number $N_w$ as labeled.
	}
	\label{f:viz_multisrc}
\end{figure*}

The pressure $p$ is shown in figure~\ref{f:viz_multisrc} for varying configurations.
At long times a spatial pattern in $p$ emerges, which changes qualitatively as the waves enter the annular region and the net interior.
For $N_w = 8$ and $12$, the pressure contours are circular at the bubble-net center, though for $N_w = 4$ no such pattern is seen.
Further, as $f$ increases, the bubble net increasingly shields the interior from the acoustic sources and the interior acoustic waves diminish in magnitude; for $f = \unit{4}{\kilo\hertz}$ a pattern cannot be discerned.
For increasing $N_w$ the amplitude of $p$ generally increases, though the penetration of the waves are most closely coupled to their frequency as the $N_w = 12$ and $f = \unit{4}{\kilo\hertz}$ case still does not noticeably penetrate the bubble net.

\begin{figure*}
	\centering
    \includegraphics[scale=1]{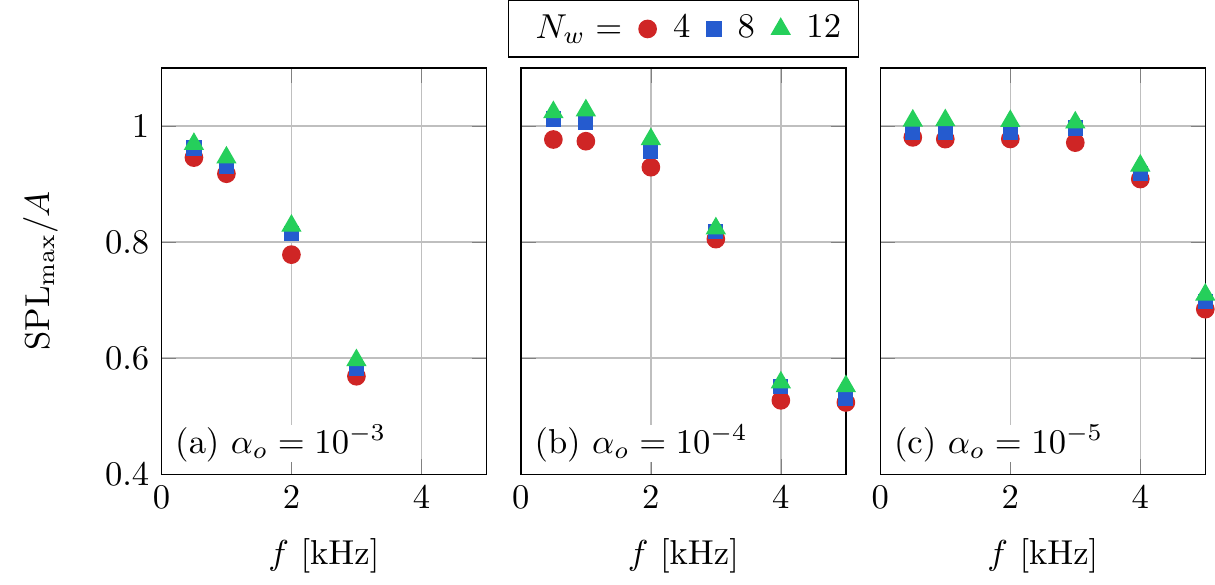}
	\caption{
	Maximum sound pressure level ($\SPL$) within the bubble-net interior 
	for varying $f$ and $N_w$, net void fractions 
    (a)--(c) $\alpha_o$ as labeled. 
	 }
	\label{f:numwhales}
\end{figure*}

Figure~\ref{f:numwhales} shows the loudness of the bubble net interior for varying vocalization frequencies, number of sources, and net void fractions.
Here, $\SPL$ is computed as the usual maximum sound pressure level (SPL) within the bubble net interior after $t = 50 \delta/c_l$ (after which they vary $< 1\%$).
Consistent with the visualizations of figure~\ref{f:viz_multisrc}, the interior loudness generally decreases with increasing $f$ and decreasing $N_w$.
For $\alpha_o = 10^{-4}$, $\SPL$ decreases with increasing $f$ until $f = \unit{4}{\kilo\hertz}$, after which it is nearly constant.
For the more dilute $\alpha_o = 10^{-5}$ cases, $\SPL$ is nearly constant with increasing $f$ until $f = \unit{4}{\kilo\hertz}$, at which point it decreases significantly.
For the less dilute $\alpha_o = 10^{-3}$, the loudness decreases a similar degree to the $\alpha_o = 10^{-4}$ cases, though for a lower $f = \unit{3}{\kilo\hertz}$.

If whales are to leverage these nets to shield and corral prey, then a small $\SPL/A$ is likely required to keep the bubble net interior as an attractive location.
If $\SPL = 0.5 A$ is used as a nominal threshold for this, then the whales must generate nets with $\alpha_o \gtrsim 10^{-4}$ to sufficiently damp their $A = \unit{180}{\deci\bel}$ vocalizations.
Nets with a void fraction this high have the additional advantage that they serve to physically trap small fish, as previously documented for bubble curtains~\citep{finley03}.
However, this configuration only makes sense when multiple whales are present, since multiple sources are required to surround the next with sound.
A seemingly more robust spiral-net configuration that is amenable to single-whale hunting is investigated next.

\subsection{Spiral-shaped bubble nets}\label{s:spirals}

As discussed in section~\ref{s:intro}, spiral-shaped bubble nets have been proposed as a possible configuration for trapping prey~\citep{leighton07}.
Such spiral nets are considered next, including variation of net void fractions and exciting acoustic frequencies.
This serves to clarify the possible acoustic mechanisms that could be present in this flow configuration.
Further, it works to resolve the parameterizations that promote robust feeding strategies..

\begin{figure*}
    \centering
    \includegraphics[scale=1]{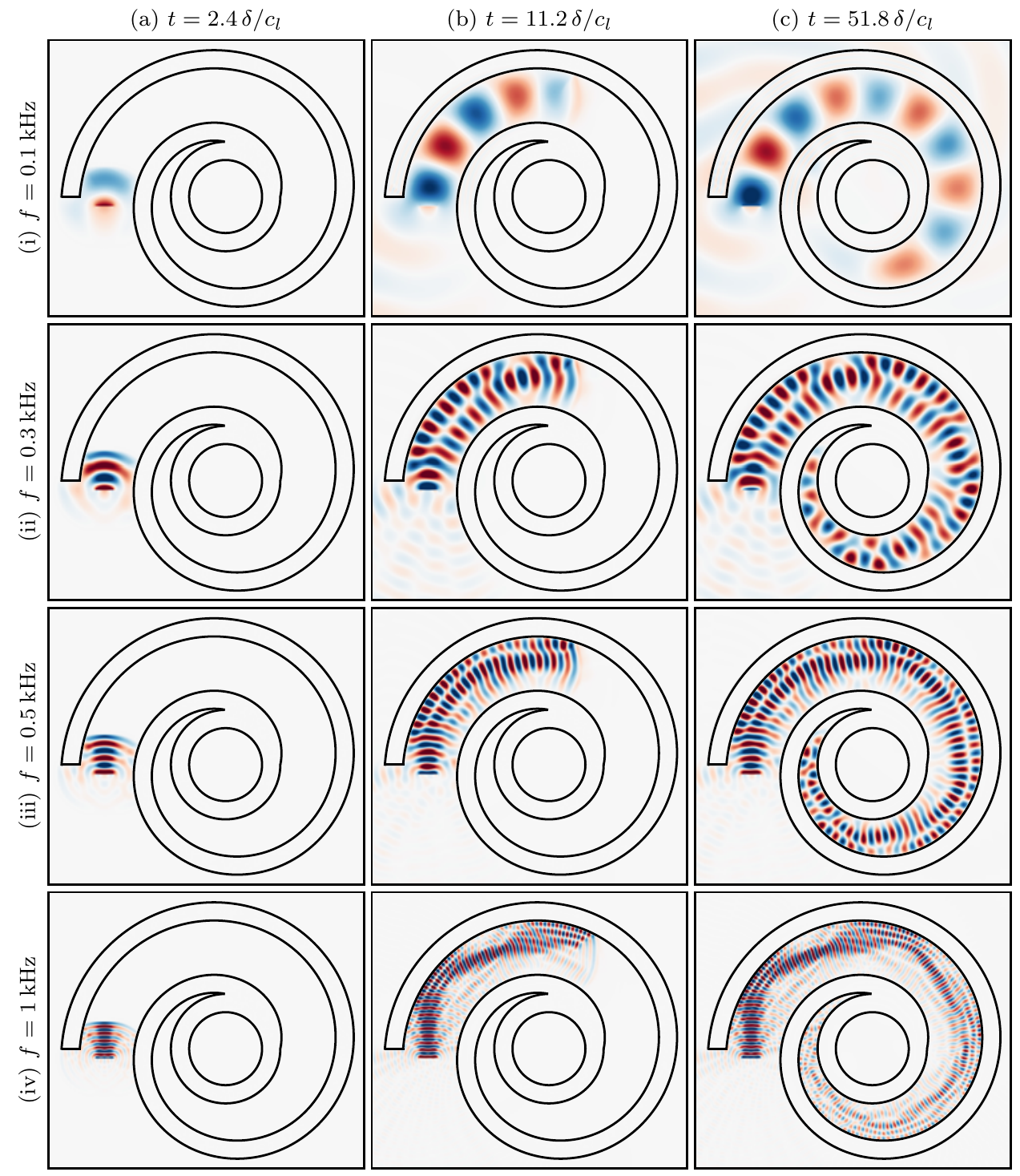}
    \caption{
        Acoustics in a $\alpha_o = 10^{-2}$ spiral bubble net for (i)--(iv) vocalization frequency $f$ at (a)--(c) times $t$.
        The normalized pressure level $p/p_\infty$ is shown using the same scale as figures~\ref{f:viz_dirsrc} and~\ref{f:viz_multisrc}.
    }
    \label{f:viz_spiral}
\end{figure*}

Figure~\ref{f:viz_spiral} shows example acoustic-spiral-bubble-net interactions for a range of vocalization frequencies and times.
For frequencies that correspond to wavelengths larger than the bubble-free-arm spacing, e.g.\ the $f = \unit{0.1}{\kilo\hertz}$ case shown here, the waves propagate around the spiral as if in a duct, with only minor attenuation due to interactions with the bubble-net wall.
For the $f = \unit{0.3}{\kilo\hertz}$ case of figure~\ref{f:viz_spiral}~(ii) a different behavior is observed.
The vocalization reflects at the bubble-wall, introducing two coherent wave patterns of similar amplitude in the constant-width portion of the bubble-free region.
This interference pattern changes in the narrowing-portion of the arm, though the wave amplitudes remain similar.
Thus, the entire bubble-free region is excited at a nearly constant amplitude.
For the $f = \unit{0.5}{\kilo\hertz}$ case a similar interference pattern is observed in the constant-width portion of the spiral, though the ever decreasing grazing angle of the reflected waves results in a quieter layer near the spiral center, consistent with the ray-tracing results of~\citet{leighton07c}.
For the $f = \unit{1}{\kilo\hertz}$ case the directionality of the acoustic waves are apparent from the first time shown, figure~\ref{f:viz_spiral}~(a,iv).
In this case the amplitude of the waves attenuates more rapidly due to reflection and transmission at the bubble-net wall.
Further, an effectively quiet route of escape from the net center exists due to the directionality of the acoustic reflections.
Of course, variations in source directions and locations could at least partially collapse this region.
However, this remains a disadvantage of the high-$f$ cases when compared to the lower $f$ cases that fully surround the center in loud vocalizations.
For these reasons, higher $f$ are not considered for this configuration.
Note that only monodisperse cases were considered for these cases. 
This is because the high impedance mismatch at the bubble-net wall results in relatively little wave transmission.
Further, polydispersity only introduced a modest effect on the refraction-dominated waveguiding observed in figure~\ref{f:viz_poly}.

\begin{figure*}
    \centering
    \includegraphics[scale=1]{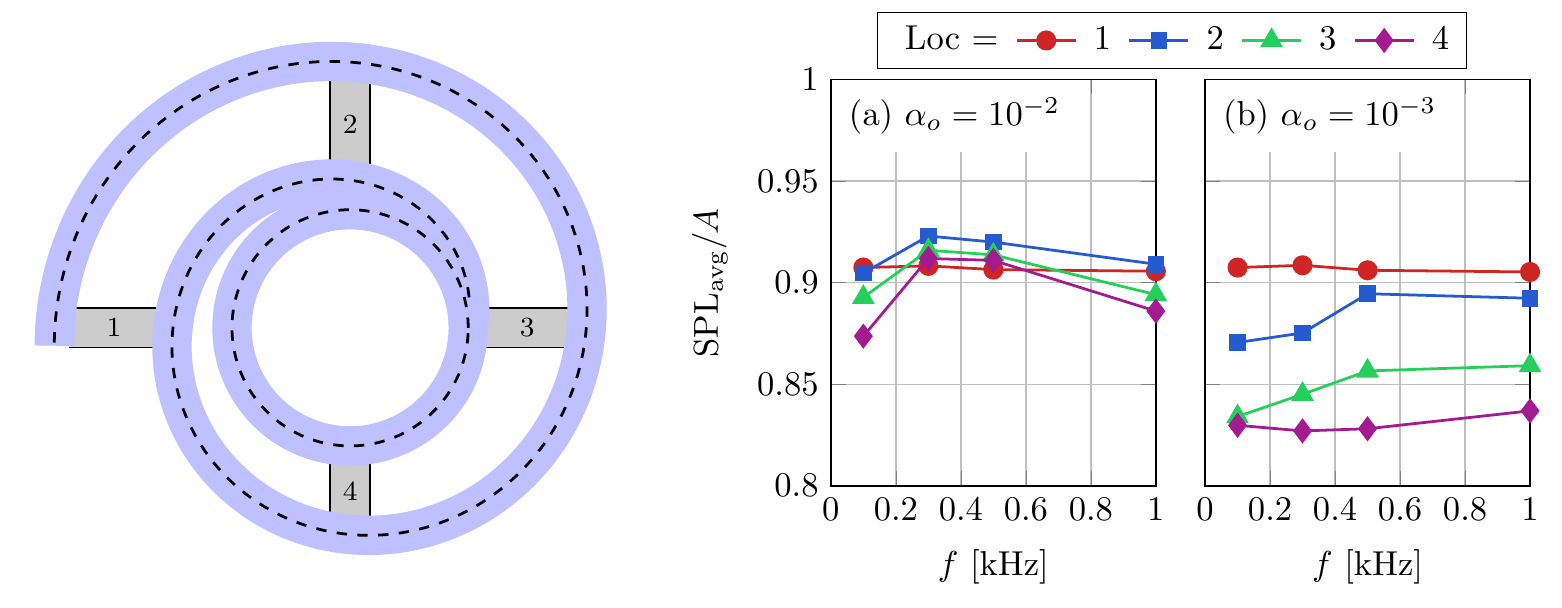}
    \caption{
        Spatially averaged long-time sound pressure level $\SPLa$ in the labeled regions 1--4 for (a) $\alpha_o = 10^{-2}$ and (b) $\alpha_o = 10^{-3}$. 
    }
    \label{f:spiral_analysis}
\end{figure*}

From the visualizations it is anticipated that $\alpha_o$ and $f$ have competing effects on the ability of the bubble net to both guide and protect the central bubble-free region from vocalizations.
These effects are first quantified in figure~\ref{f:spiral_analysis} by measuring the averaged sound pressure $\SPLa$ in four $0.2r_o$-thick patches located along the spiral portion of the net.
As expected, $\SPLa$ at location 1, prior to significant interaction with the bubble wall, is effectively independent of $f$ for both $\alpha_o$.

For the higher void-fraction case ($\alpha_o = 10^{-2}$), $\SPLa$ is similar for all 4 locations along the spiral, as there is greater internal reflection and smaller transmission due to the high impedance mismatch.
For the lower volume fraction case ($\alpha_o = 10^{-3}$), where the impedance mismatch is smaller, there is greater transmission of waves and consequently a significant reduction in $\SPLa$ along the spiral.
These results are independent of $f$, with only minor variations due to differing spatial wave patterns within the spiral over the range of frequencies considered.

\begin{figure}[ht]
	\centering
    \includegraphics[scale=1]{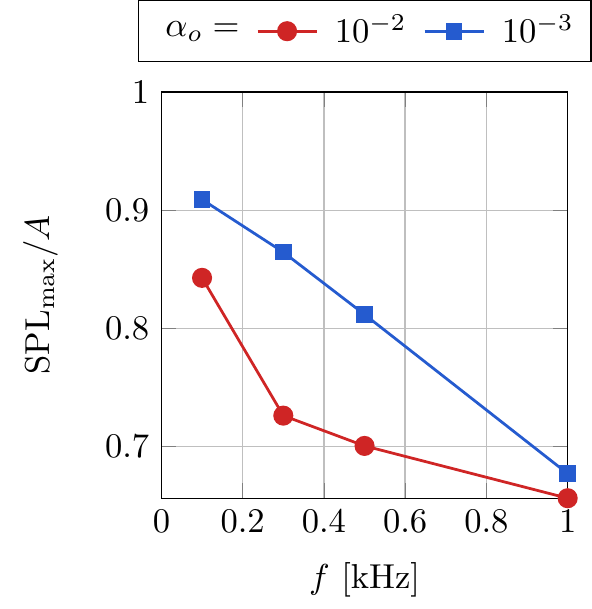}
	\caption{
        Maximum sound pressure level $\SPL$ in the bubble-net center for varying $f$ and (a) $\alpha_o = 10^{-2}$ and (b) $\alpha_o = 10^{-3}$.
    }
	\label{f:spiral_inner}
\end{figure}

The $\alpha_o = 10^{-3}$ cases were shown to allow a portion of the vocalizations to escape the bubble net entirely, which decreased their measured amplitude as they propagate around the spiral.
However, this lower $\alpha_o$ also means that the remaining waves can more easily penetrate the central bubble net region, where the prey are corralled.
Figure~\ref{f:spiral_inner} illustrates this trade-off via the maximum sound pressure level in the bubble-net center for varying $\alpha_o$ and $f$.
For $f < \unit{1}{\kilo\hertz}$, $\SPL$ is significantly smaller for the $\alpha_o = 10^{-2}$ than $10^{-3}$ cases.
Thus, even though a portion of the wave can escape the spiral net for the $\alpha_o = 10^{-3}$ cases, it still remains louder in the central bubbly-free region.
As a result, the $\alpha_o = 10^{-2}$ configurations are preferable for these $f$ in terms of both surrounding the entire net with loud vocalizations and keeping the interior quiet.
For $f = \unit{1}{\kilo\hertz}$ similar $\SPL$ are observed for both $\alpha_o$.
This is due to the trade-off between increased refraction in the bubbly arm of the spiral and the inability of these higher frequencies to penetrate the central net annulus.
This behavior is expected for still higher $f$ due to the directionality of the waves in this regime.

Thus, for low-frequency $f \lesssim \unit{1}{\kilo\hertz}$ vocalizations, a higher $\alpha_o \gtrsim 10^{-2}$ void fraction is preferable in order to keep the waves in the bubble-net spiral. 
For higher frequencies $f \gtrsim \unit{1}{\kilo\hertz}$, $\SPL$ is relatively insensitive to $\alpha_o$ and thus this parameter is less important. 
However, as was shown in figure~\ref{f:viz_spiral}, the directionality of these high frequency vocalizations means that a quiet region can form at the interior layer of the spiral, leaving the prey a possible escape route.

\section{Discussion and conclusions}\label{s:conclusion}

As a step towards fully understanding the complex humpback whale bubble-net feeding process three possible acoustic mechanisms were assessed: wave-guidance and bubble-wall shielding for annular nets and vocalization guiding for spiral configurations.
For this, a fully-coupled compressible bubbly flow model capable of representing the modified mixture speed of sound, geometric effects of the curved bubble wall, and nonlinear and collective bubble dynamics was used.
This model was solved using a high-order interface capturing scheme that minimized spurious oscillations near material interfaces.
A configuration of acoustic-generating sources and bubbly regions was used as as a model of actual vocalizing humpback whales and the bubble nets they generate.
Simulations of this system were analyzed and connected to observed whale feeding phenomena.

A wave-grazing flow configuration was considered to determine if waveguiding in the bubbly region could efficiently keep the bubbly region loud.
Analysis showed that for the parameterizations considered only modest waveguiding could be observed.
This was most prominent for relatively low frequencies ($f = \unit{1}{\kilo\hertz}$), but still only encompassed less than half of the bubble net.
To ensure that this was not due to the model parameterization, this conclusion was shown to be unchanged when considering bubble population polydispersity, varying bubble-net thickness, acoustic frequencies, and their directionality, breadth, and amplitude.

A configuration representing multiple vocalizing whales was used to quantify the attenuation and acoustics of an annular bubble-net and the region that surrounded it.
Qualitatively different acoustic patterns were observed, depending upon the number of whales present.
The degree of attenuation was most strongly dependent on the frequency of the vocalizations, with significant attenuation of the $\unit{180}{\deci\bel}$ waves down to about $\unit{90}{\deci\bel}$ for a range of cases that overlapped with possible whale vocalization frequencies and net void fractions.
This suggests that it is possible humpback whales utilize these bubbly regions as a shield, but only if multiple whales are cooperating.

The weakness of the acoustic shielding hypothesis is associated with the required number of cooperating whales to utilize it.
A previously proposed spiral bubble net configuration that could be utilized by a single whale was also considered~\citep{leighton07}.
Indeed, observations suggest that the shape might be closer to spiral than annular.
With a single model whale vocalizing into the bubble-free end of the spiral net, the guidance of the acoustic waves through the spiral was observed.
However, their behavior depended upon the vocalization frequency and net void fraction.
For example, vocalizations of wavelengths near the width of the bubble-free arm were simply guided through this region, without noticeable reflection at the bubble walls, whereas higher frequency cases displayed varying degrees of reflections and, thus, loudness as they propagated through the spiral.
Importantly, for nets with smaller void fractions, low frequency vocalizations were able to penetrate the bubble arm, reducing their magnitude when they reach the central bubble net region.
For higher frequencies, both reflection and refraction and the bubble-net walls resulted in a directional acoustic behavior at ever decreasing grazing angles.
These cases kept the most of the bubble-free region loud, though for sufficiently high frequencies a quiet region for the model prey to escape exist.
This set of competing effects was, in part, quantified by the maximum sound pressure level observed in the bubble-net center.
This metric suggested that the higher void-fraction $\alpha_o = 10^{-2}$ nets considered were superior to the $\alpha_o = 10^{-3}$ cases, even though they guided the entirety of the whale vocalizations towards the bubble-net center.

Additional field observations are required to further clarify the space of possible configurations.  
For example, the exact spatial locations and directions of the whales during feeding would better illuminate their behaviors.
A better estimation of the expected bubble void fractions in the nets could also possibly rule out acoustic mechanisms or vocalization frequencies, since this parameter was intimately related to the viability of the configurations considered.

\begin{acknowledgments}

We thank Dr.\ Kevin Schmidmayer for numerous fruitful discussions.  
This work was supported by the Office of Naval Research under grant number N0014-17-1-2676.
Associated computations utilized the Extreme Science and Engineering Discovery Environment, which were supported by the US National Science Foundation under grant number~TG-CTS120005.

\end{acknowledgments}


\end{document}